\documentclass[fleqn,12pt]{wlscirep}
\usepackage{setspace}
\usepackage{subcaption}
\usepackage{caption}
\usepackage{color}
\usepackage[symbol]{footmisc}
\usepackage{lineno}
\usepackage{amsmath}
\usepackage{multirow}
\usepackage{adjustbox}
\usepackage{float}
\usepackage{threeparttable}
\usepackage{aas_macros}

\title{Sub-second periodic radio oscillations in a microquasar }

\author[1,2,14] {Pengfu Tian}
\author[1,2,14] {Ping Zhang}
\author[1,2,14*] {Wei Wang}
\author[3,13,14] {Pei Wang}
\author[4] {Xiaohui Sun}
\author[2,3] {Jifeng Liu}
\author[5,6*] {Bing Zhang}
\author[1,7] {Zigao Dai}
\author[8] {Feng Yuan}
\author[9] {Shuangnan Zhang}
\author[10] {Qingzhong Liu}
\author[3,11] {Peng Jiang}
\author[10] {Xuefeng Wu}
\author[3] {Zheng Zheng}
\author[1,2] {Jiashi Chen}
\author[3] {Di Li}
\author[1,11] {Zonghong Zhu}
\author[3,12] {Zhichen Pan}
\author[3,12] {Hengqian Gan}
\author[1,2] {Xiao Chen}
\author[1,2] {Na Sai}

\affil[1]{Department of Astronomy, School of Physics and Technology, Wuhan University, Wuhan 430072, People's Republic of China;}
\affil[2]{WHU-NAOC Joint Center for Astronomy, Wuhan University, Wuhan 430072, People's Republic of China;}
\affil[3]{National Astronomical Observatories, Chinese Academy of Sciences, Beijing 100012, People's Republic of China; }
\affil[4]{School of Physics and Astronomy, Yunan University, Kunming 650500, People's Republic of China;}
\affil[5] {Nevada Center for Astrophysics, University of Nevada, Las Vegas, NV, USA;}
\affil[6] {Department of Physics and Astronomy, University of Nevada, Las Vegas, NV, USA;}
\affil[7]{School of Astronomy and Space Science, University of Science and Technology of China, Hefei 230026, People's Republic of China;}
\affil[8]{Shanghai Astronomical Observatory, Chinese Academy of Sciences, 80 Nandan Road, Shanghai 200030, People's Republic of China;}
\affil[9]{Key Laboratory of Particle Astrophysics, Institute of High Energy Physics, Chinese Academy of Sciences, Beijing 100049, People's Republic of China;}
\affil[10]{Purple Mountain Observatory, Chinese Academy of Sciences, Nanjing 210023, People's Republic of China;}
\affil[11]{Henan Academy of Sciences, Zhengzhou 450046, Henan, People's Republic of China;}
\affil[12] {Guizhou Radio Astronomy Observatory, Guizhou University, Guiyang, 550025, People's Republic of China.}
\affil[13]{Institute for Frontiers in Astronomy and Astrophysics, Beijing Normal University, Beijing 102206, People's Republic of China}
\affil[14] {These authors contributed to this work equally.}
\affil[*] {Emails: wangwei2017@whu.edu.cn,bing.zhang@unlv.edu}
\begin{abstract}
{\bf Powerful relativistic jets are one of the ubiquitous features of accreting black holes in all scales\cite{mirabel1999,Remill2006,zensus1997}. GRS 1915+105 is a well-known fast-spinning black-hole X-ray binary \cite{mcclint2006} with a relativistic jet, termed as a ``microquasar'', as indicated by its superluminal motion of radio emission \cite{mirabel1994,Fender2004}. It exhibits persistent x-ray activity over the last 30 years, with quasi-periodic oscillations of $\sim 1-10$ Hz\cite{belloni2000,misra2020,zhang2020} and 34 and 67 Hz in the x-ray band \cite{belloni2013}. These oscillations likely originate in the inner accretion disk, but other origins have been considered \cite{ingram2019}. Radio observations found variable light curves with quasi-periodic flares or oscillations with periods of $\sim 20-50$ minutes \cite{pooley1997,rodri1997,klein2002}. Here we report two instances of $\sim$5 Hz transient periodic oscillation features from the source detected in the 1.05-1.45 GHz radio band that occurred in January 2021 and June 2022, respectively. Circular polarization was also observed during the oscillation phase.
}
\end{abstract}

\begin{document}
\maketitle

We used the Five-hundred-meter Aperture Spherical radio Telescope (FAST) \cite{jiang2020} to perform a high-sensitivity, millisecond-time-resolution study of GRS 1915+105, aiming to study the fine details of jet dynamics. We performed tracking-mode observations on the source in the 1.05-- 1.45 GHz band with the central beam of the 19-beam receiver and a 49.152 microsecond sample time starting on January 25 2021 01:35:00 (UTC). The continuous observations lasted ninety minutes, with full Stokes polarization parameters recorded. After data reduction and calibration (see details in Methods), we derived the variations of the total intensity, degrees of linear polarization (LP) and circular polarization (CP), and linear polarization position angle (PA) over the observing time intervals with a time resolution of $\sim$ 0.002 s, as presented in Figure \ref{fig:lc-pol}.

To study the variation properties of radio emission, the dynamical power spectrum of the light curve of radio flux density is calculated and displayed in the bottom panel of Figure \ref{fig:lc-pol}. The power spectrum over time shows two striped structures appearing in the middle interval of the observations, indicating transient quasi-periodic oscillations (QPOs) at $\sim 0.196\pm 0.002$ s and $0.104\pm 0.003$ s in the radio light curve (all uncertainties are at the 1$\sigma$ level). The two vertical dashed lines divide the observing time into three time domains: Epoch A is when the total intensity flux density was in a relatively stable state at $\sim 400-420$ mJy before the appearance of the QPOs. The radio spectral index $\alpha$ in the 1.05 -- 1.45 GHz band varies from $\sim -0.6$ to $-0.5$; Epoch B is the period of periodic oscillations, which lasted for about 1260 seconds. During the epoch, the flux density increased from $\sim 450$ mJy to $\sim 660$ mJy, and $\alpha$ evolved from $\sim -0.5$ to $-0.35$; Epoch C is the epoch after the periodic oscillations disappeared. The flux density reached a stable level of $\sim 660$ mJy with $\alpha\sim -0.3$. One can see that the transient sub-second periodic oscillations only occurred during the period of a rapid increase of radio flux density. The significant change of $\alpha$ suggests the occurrence of an ejection event \cite{fenderBG2004,Fender2004}.

Significant polarization evolution with flux and time was also observed. In Figure \ref{fig:POL_DIS}, we present the distributions of linear polarization, circular polarization and position angle in three epochs defined above. The linear polarization degree was distributed normally around 26.5$\%$ in Epoch A, increased to $\sim 28.5\%$ in Epoch B, and continued to increase to $\sim 30.5\%$ in Epoch C. The circular polarization evolution shows a different evolution behaviour with time. A clear double-peak symmetrical distribution was observed before the QPO phase (Epoch A) with both left and right CPs. The distribution became a single peak with pure left CP in Epoch B, and switched back to the double peak distribution again in Epoch C with a little bit more right CP than left CP. The linear position angle has the similar distributions for both Epochs A and C with the peak value of $\sim 87.5^\circ$. During Epoch B, the position angle became systematically larger with the peak distribution around $PA\sim 88.5^\circ$.

The transient sub-second quasi-periodic oscillations have the complicated temporal variation structure which could be connected to jet dynamics of the black hole (BH). We then perform a wavelet analysis of the light curves during the epoch of periodic oscillations (see Methods). The wavelet analysis result of the power profiles of the radio light curve with the QPO signal is presented in the top panel of Fig. \ref{fig:wavefold}. The contours of wavelets show the variation characteristics of the QPOs in both frequency and time domains with the 95\% confidence level for the global wavelet spectrum. The local wavelet power spectrum example has a duration of 20 s, which displays a significant 5-Hz signal across essentially the entire time span. The 10-Hz harmonic signal is weaker and more sparse. In the high time-resolution wavelet power diagram the 5-Hz signal still shows a discontinuous behavior, with the QPO signal disappearing sometimes. A statistical study of the timescales of the detected transient QPO signals finds the duration of the 5-Hz QPO to be about $0.4 - 12$ s with a peak of $\sim 0.7$ s. The typical duration of the 5-Hz QPO signal at $\tau \sim 0.7$ s defines a characteristic scale $\sim \tau c \sim 2\times 10^{10}$ cm, $c$ is the light speed, which may be related to the typical size of the QPO emission region or lower limit of the emission height.

To probe the variation characteristics of the QPOs, we fold the radio light curves at the period of 0.196 seconds for the Epoch B data. In the bottom panel of Fig. \ref{fig:wavefold}, the QPO pulse profiles for the 5-Hz signals are presented. The pulse profile shows a broad single peak covering more than 70\% of the whole phase. The pulse-folding technique also constrains a dispersion measure (DM) of $\sim 230 - 280$ pc cm$^{-3}$ for GRS 1915+105 (see Methods). Additionally, we also fold the light curves of three polarization parameters (LP, CP, PA) at the same period during Epoch B, showing variations of these polarization parameters over the QPO pulse profile. In particular, the circular polarization shows a variation pattern similar to the QPO profile.

GRS 1915+105 has shown variable linear and circular polarization of radio emission by relativistic electrons which reveals a large-scale magnetic field structure in the outflow \cite{fender2002}. Circular polarization of $\sim -1\%$ was detected during the QPO epoch. In a relativistic jet, there may be two physical mechanisms for CP: an intrinsic CP from synchrotron radiation and LP to CP conversion (repolarization by Faraday conversion)\cite{fender2002,beck2002}. A magnetic field configuration perpendicular to the line of sight, would induce Faraday conversion, especially with the presence of copious low-energy electrons\cite{beck2002}.
If the large-scale magnetic field along the line of sight changes orientation with time, it also induces modulation of the observed CP.

The detection of sub-second periodic oscillations in the radio band brings the first direct evidence of sub-second QPOs within the jet emission region in a stellar-mass BH system. To confirm the genuine origin of the QPO, we searched for similar sub-second QPO features in GRS 1915+105 in the archived data observed with FAST from $2020 - 2022$. We detected another transient QPO signature at $\sim 0.21\pm 0.02$ s which lasted $\sim 80$ seconds with the observations performed on 2022-06-16  (see Fig. \ref{fig:event2022}), with the 19-beam receiver recording the data of all beams simultaneously. During the observations, the central beam (M01, field of view (FOV) of 3 arcmin) was toward the target GRS 1915+105, while the beams M02 -- M19 were toward other sky regions for background monitoring. We applied the Fast Fourier Transform (FFT) analysis to all lightcurves including beam M01 on GRS 1915+105, and other beams on the off-source sky regions in the same time interval (see Fig. \ref{fig:event2022}). Only the M01 data showed the QPO peak feature at around 5 Hz in the power spectrum. Other beam light curves only showed  fluctuations in the power spectra with no QPO features detected. This strongly suggested that the sub-second QPOs at $\sim 5$ Hz was again from GRS 1915+105. This observation confirmed the QPO signature detected in January 2021, suggesting that the $\sim$ 5 Hz frequency is representative for this microquasar system. During the second event, the radio flux was relatively steady at a level around 350 mJy, with the measured LP $\sim 6.5\%$, CP $\sim -1.3\%$ and PA $\sim 96^\circ$. The spectral index $\alpha$ evolved from $-0.08$ to $\sim -0.01$ during the period.

Low frequency quasi-periodic oscillations have been detected in black-hole X-ray binaries in the optical, infrared, ultraviolet and X-ray bands \cite{ingram2019,Kalamkar2016}. These QPOs likely originate from the inner accretion disk, but other origins from the corona or the outflow are also possible\cite{ingram2019}. Models invoking different physical mechanisms have been proposed to interpret these QPOs: e.g., accretion ejection instability \cite{tagger1999}, propagating oscillatory shocks in the disk \cite{chakra2008}, relativistic precession of the inner accretion flow \cite{ingram2009} or jet base \cite{stevens2016}. Radio emission directly probes jet emission. Thus, the first detection of sub-second modulations in radio emission provides an unambiguous connection between the QPOs and the dynamics of the jet. The new phenomenon could arise from the precession of a magnetised relativistic jet with a warped accretion disk \cite{miller2019,ma2021}. However, it is not clear whether this variability can survive to the optically thin region of the jet, and detailed modelling is needed to test this scenario. While it is interesting to speculate on a connection between disk oscillations and jet oscillations, at present there is no confirmed mechanism to propagate oscillations from the disk to the jet.

Nowadays GRS 1915+105 appears in a dimming X-ray state. During our FAST observations both in 2021 and 2022, GRS 1915+105 did not show any enhancement of emission in X-rays, suggesting possible obscuration of the X-rays (see long-term X-ray light curves and discussion on the special state in Methods). In general, the launch of the jet may require a special condition at the engine, and a rapidly spinning black hole with a magnetized, varying accretion flow may be in operation in GRS 1915+105 to power these relativistic jets.

\renewcommand{\figurename}{Fig.}
\renewcommand{\thefigure}{\arabic{figure}$ $}

\begin{figure}
\centering
\includegraphics[width=.6\textwidth]{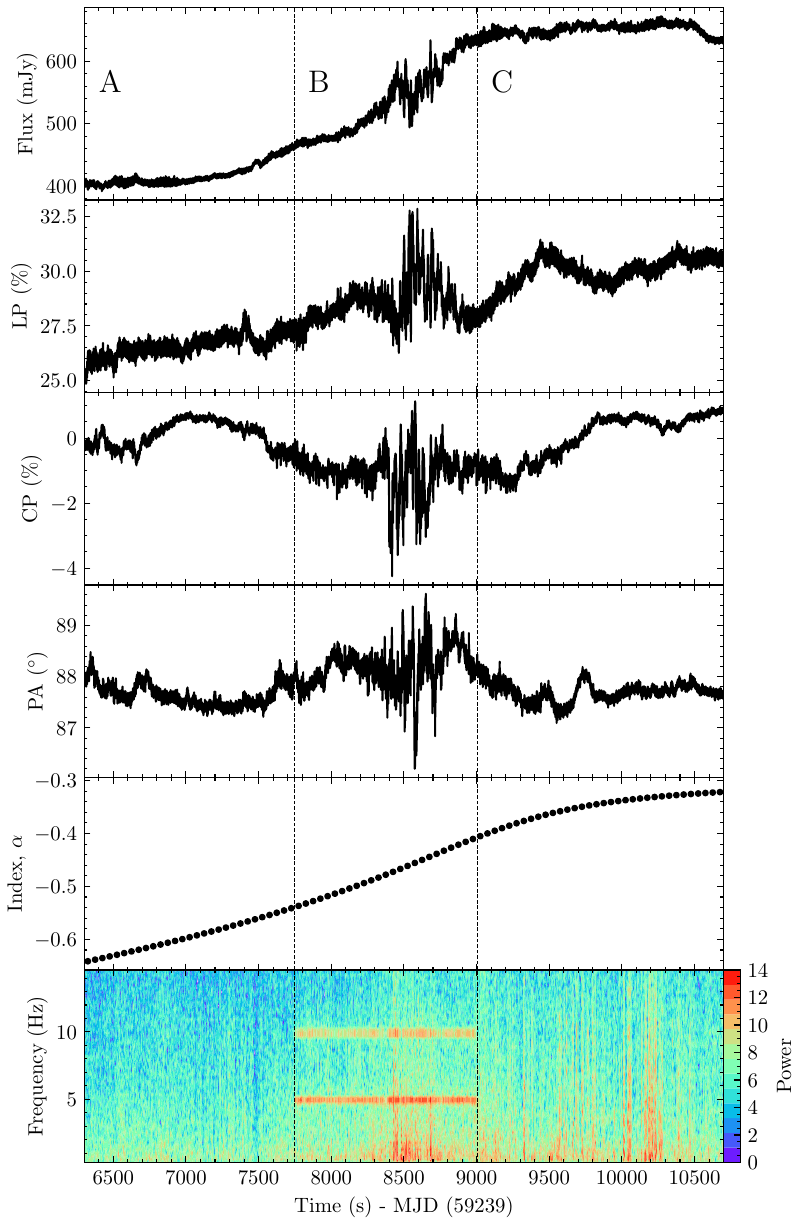}
\caption{{\bf Light curves during the QPO phase in 2021.} Evolution of total intensity flux density, degree of linear polarization (LP), degree of circular polarization (CP), linear polarization position angle (PA), spectral index and dynamic power density spectrum (PDS) with observations from 2021-01-25:01:35:00 to 2021-01-25:03:04:57 (UTC). The two dashed lines divide the time domain into three regimes: (Epoch A) flux density around 400-420 mJy; (Epoch B) flux increased from $\sim$ 450 mJy to 660 mJy, the 5 Hz and 10 Hz QPOs appeared with a duration of $\sim$ 1260 seconds; (Epoch C) flux shows a plateau of $\sim 660$ mJy and the QPOs disappeared. Linear polarization has a slow rising from 25$\%$ to 31$\%$ as the flux density rising, while circular polarization evolves from around zero in Epoch A, to $\sim -(1-2)\%$ in Epoch B then returns to zero in the end of the observation. Position angle is determined to be around 87.5 degrees in Epochs A and C, and $PA\sim 88.5^\circ$ in Epoch B. The evolution of spectral index from $\sim -0.65$ towards -0.3 indicates the system evolving from optically thin to the optically thick. In the bottom panel, we calculated the dynamical power spectrum with combining PDS for each data set of 4 second duration. The QPO signals at $\sim$ 5 and 10 Hz are only detected during Epoch B.}
\label{fig:lc-pol}
\end{figure}

\begin{figure}
\centering
\includegraphics[width=.6\textwidth]{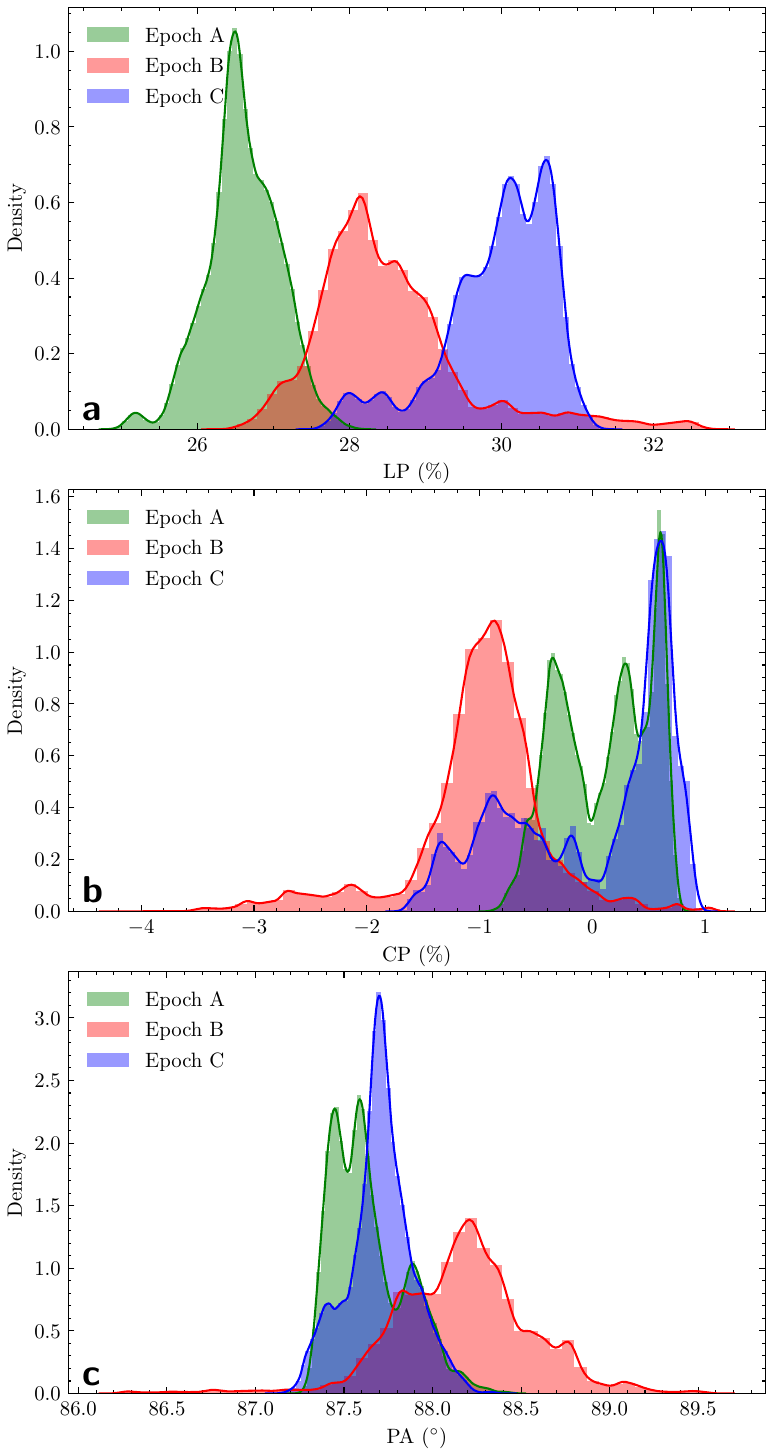}
\caption{{\bf Evolution of the polarization parameters.} \textbf{a-c, }The value distributions of the linear (\textbf{a}), circular polarization (\textbf{b}) and position angle (\textbf{c}) in three observational Epochs A, B and C. The solid lines are the KDE (Kernel Density Estimation) smoothing curves. Linear polarization rises slowly with the time , while circular polarization and position angle show different behaviors between Epoch B (the QPO regime) and Epochs A, C (see text for detailed description). }
\label{fig:POL_DIS}
\end{figure}

\begin{figure}
\centering
\includegraphics[width=.6\textwidth]{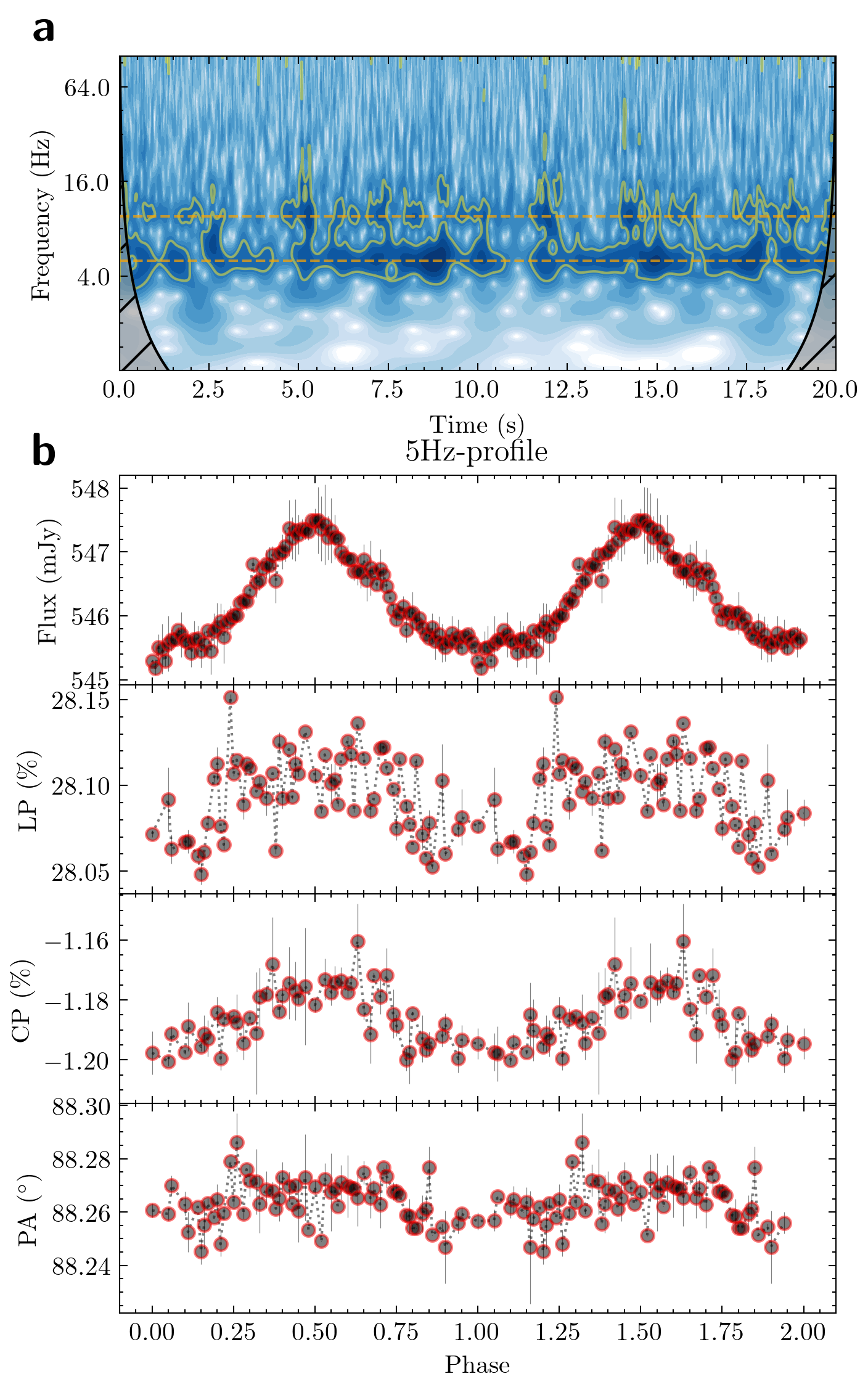}
\caption{{\bf Fast variations of the QPO.} {\bf a,} The wavelet analysis result of the flux for a time interval of 20 seconds as an example, the contour plots show the detected periodic signals and their evolution with time. The orange dashed lines annotate the centroid frequencies of the periodic signals. The discontinuous and scattering features of QPOs indicate that the signals are not detected sometimes and may not appear simultaneously. {\bf b,} from top to bottom, the pulse profiles of 5 Hz regime over QPO phases folded at 0.2 second period for flux (Stokes I), linear polarization (LP), circular polarization (CP) and position angle (PA). The pulse profiles of the 5 Hz QPO are Lorentzian-like, showing a single peak. In addition, the polarization profiles, specially CP, show similar modulations in phase with the flux profile (amplitudes variations of $\sim 2-3\%$). The error bars are given with the ranges of 1 $\sigma$. }
\label{fig:wavefold}
\end{figure}

\begin{figure}
\centering
\includegraphics[width=.95\textwidth]{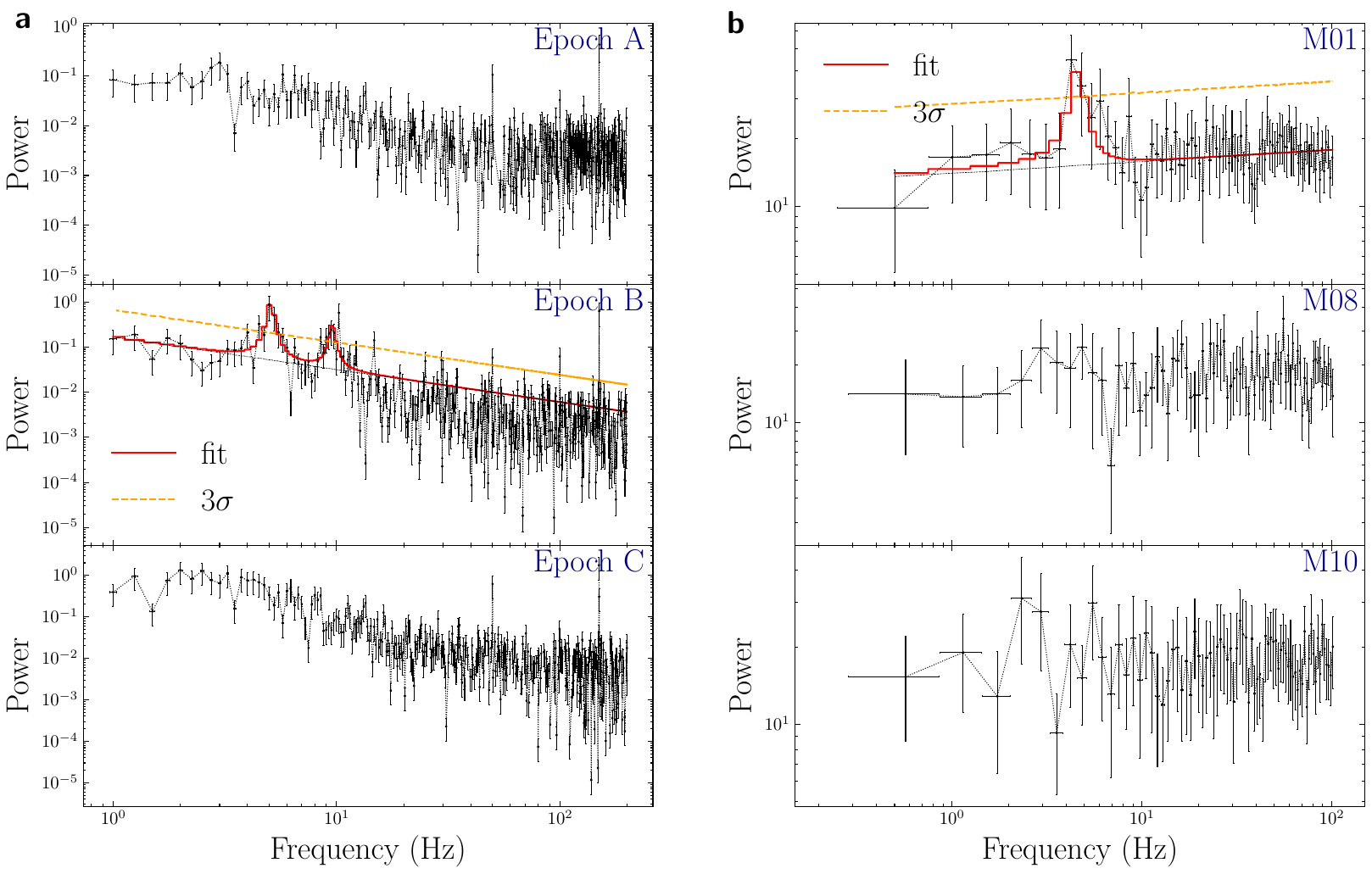}
\caption{{\bf The power spectra of radio light curves based on the FAST observational data.} {\bf a,} The power density spectra (PDSs) of three light curves selected from the three epochs observed on 2022-01-25 (A, B and C defined in Fig. \ref{fig:lc-pol}). The PDS in Epoch B is fitted with a power-law component and two Lorentzian functions. The colored dashed line in the middle panel indicates the confidence level at $3\sigma$ (the details of uncertainty calculation are described in Methods). {\bf b,} The PDSs of the radio light curves observed on 2022-06-16 simultaneously recorded by the central beam M01 toward GRS 1915+105 and the other beams, e.g., M08, M10, toward other sky regions. The sub-second QPOs around 5 Hz were only reported from the target source. The power spectrum from M01 is fitted with a power-law component and a Lorentzian function. The colored dashed line in the top panel indicates a confidence level at 3$\sigma$.}
\label{fig:event2022}
\end{figure}

\newpage
\section*{Methods}
\setcounter{figure}{0}
\renewcommand{\figurename}{Extended Fig.}

\section{Observations}
\begin{figure}
\centering
\includegraphics[width=.7\textwidth]{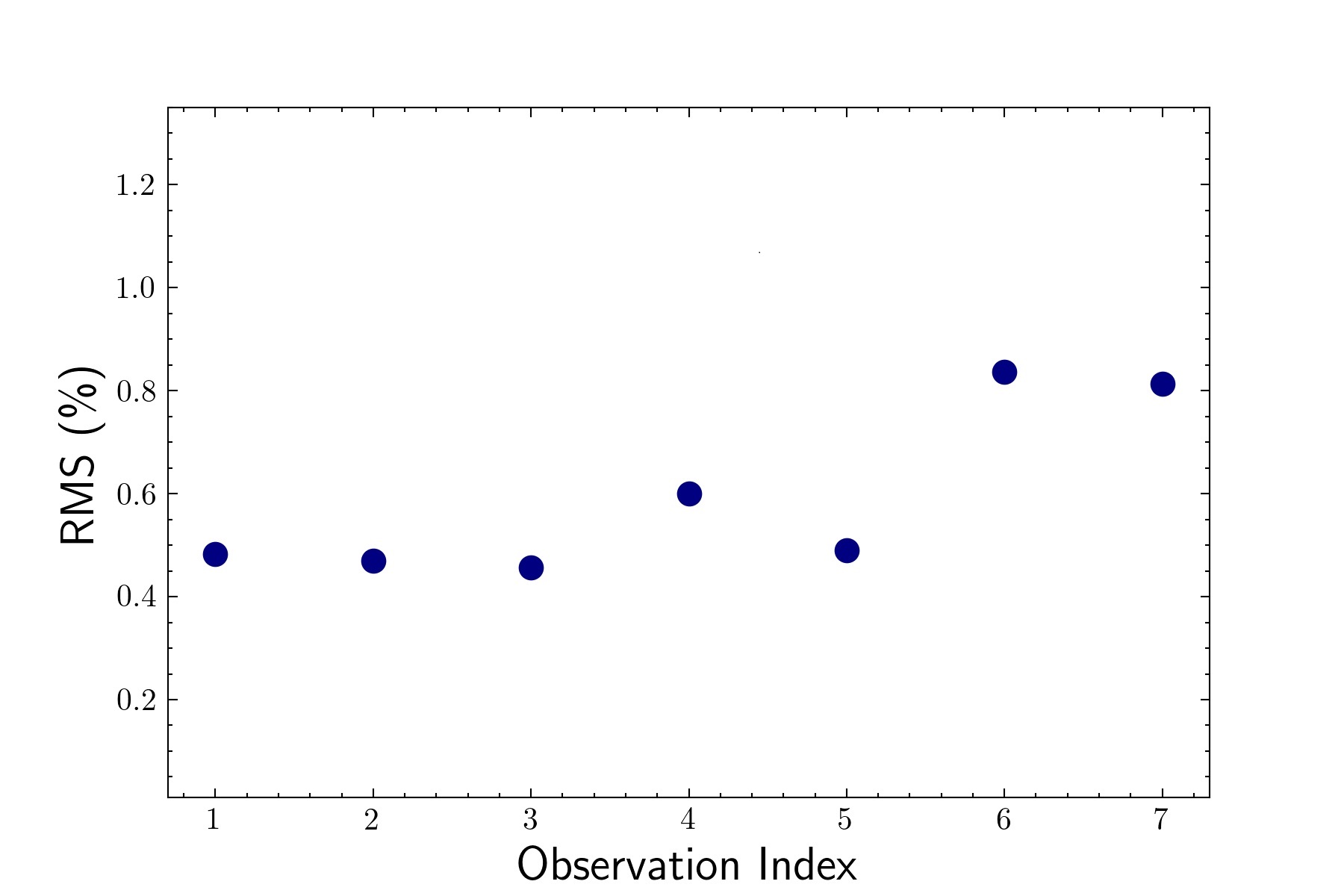}
\caption{{\bf Stability of FAST performance.} The rms of the flux when the feed source is pointing to the background sky during our FAST observations.}
\label{fig:offsourcerms}
\end{figure}

The Five-hundred-meter Aperture Spherical radio Telescope (FAST) is the largest single dish and the most sensitive radio telescope of the world in 0.07-3 GHz frequency range. The frequency coverage of the 19-beam receiver is 1.05-1.45 GHz. The system temperature is a function of zenith angle and can be fitted with a modified arctan function validly no more than 40 degrees from the zenith, because the background noises of this observation agree with the average level of the long-term noise monitoring (\textcolor{blue}{Extended Fig.} \ref{fig:offsourcerms}). To facilitate the calibration, reference signals produced by the  noise diode are injected into the receiving system. The temperature of the reference signal is about 1.1 K for low power mode and 12.5 K for high power mode \cite{jiang2020}, respectively. The measured temperature uncertainty of diode is $\sim$ 1\%, which would lead to $\sim$ 2\% accuracy in flux calibration.

Here, we have carried out a FAST observation on the microquasar GRS 1915+105 in tracking mode at 1.05-1.45 GHz band with the central beam of the 19-beam receiver \cite{jiang2020}, starting at January 25 2021 01:35:00 (UTC) with a 49.152 microsecond sample time in a duration of 90 minutes. The resolution of the central beam is $\sim 2.9'$. Before and after the tracking mode observations, the pattern of the on-off mode in which the noise diode was continuously switched on and off has been performed for two time intervals: from 01:25:00 - 01:30:00 (UTC) and from 03:10:00 - 03:15:00 (UTC), which are used for calibration processes. This led to the detection of the first QPO event.

A second event was also detected by FAST during an observation performed on June 16 2022 from 17:42:40 (UTC) to 17:47:30 (UTC) in the tracking mode at 1.05-1.45 GHz band with the central beam of the 19-beam receiver. The data from all 19 beams were recorded for this observation. The central beam M01 was beamed towards the source, and the other 18 beams (M02 -- M19) were beamed towards the off-source sky regions. Before and after the tracking mode observations, the pattern of the on-off mode in which the noise diode was continuously switched on and off has been performed.

\section{Extracting light curves of radio flux and polarization}
Here we show the method details to reduce data and calibrate the total intensity and polarization. The data of FAST are recorded in {\tt PSRFITS} format \cite{hotan2004}. Firstly, we use {\tt astropy} package \cite{astropy2018} to do the preprocessing for {\tt FITS} data files. For each of {\tt FITS} file, we do the re-sampling for the original data and extract the frequency band and time from 4096 frequency channels and 128 subints, and then combine the re-sampled preprocessed data files. The {\tt PRESTO} \cite{ransom2011} would produce a time series which is the uncalibrated lightcurve from the combined file. Meanwhile, {\tt PRESTO} can find the radio frequency interferences (RFI) and create a mask to eliminate these narrow-band signal-like noise. The noisy broadband signal in periodogram may be caused by RFIs, e.g., the variation of the feed source forms a 0.1 Hz noise, the frequency of alternating current in electronic system may also cause a noise of 50 Hz and the harmonic component in the whole observation time.

\begin{figure}
\centering
\includegraphics[width=1.\textwidth]{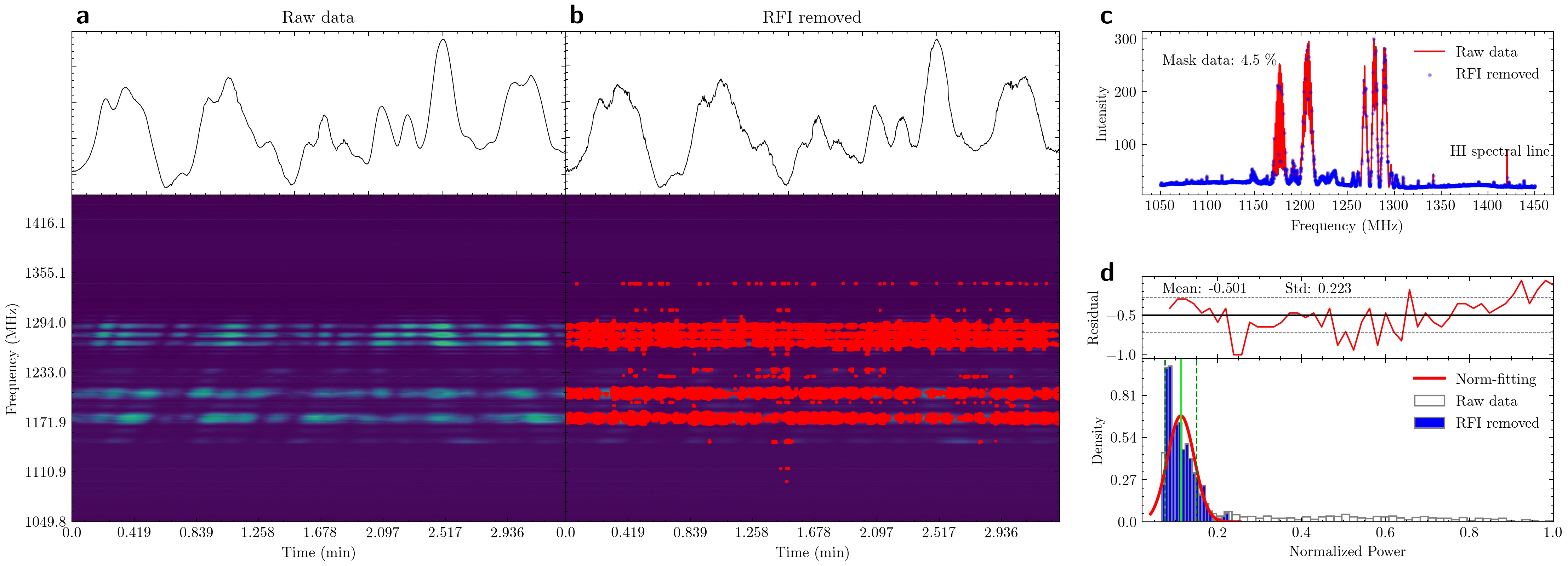}
\caption{{\bf A demonstration of RFI-mitigation experiment using the two-dimensional wavelet algorithm.} \textbf{a,} signal intensity as a function of frequency and time (waterfall) plot of the raw data of GRS 1915+105 between MJD 59239.09766 - 59239.09997 (bottom subplot) and frequency-averaged light curve (upper subplot). \textbf{b,} the red dots at the waterfall plot represent the masked RFI contaminated data by using the two-dimensional wavelet algorithm and then fill these masked data by the median values. \textbf{c,} comparison of frequency bandpass for the raw data and RFI removal result. \textcolor{blue}{\textbf{d,}} comparison of histogram for raw data (white) and RFI removed data (blue, the red line is the gaussian fitting with the value of Chi-square is less than 5$\%$).}
\label{fig:RFIs}
\end{figure}

{\bf RFI removing processes} \\
We used the two-dimensional wavelet transform method to mask the RFI contaminated data and then fill these masked data by the median values. The two-dimensional wavelet transform can be used to extract time-frequency structural features of RFIs from the noise data along the horizontal, vertical and diagonal lines of the feature matrices, respectively, i.e., the narrow-band frequency domain RFI and the border-band impulsive RFI can be extracted in the horizontal and vertical directions, while the diagonal features indicate the isolate abnormal values. We project anomalous signals into these three feature dimensions, then smooth the obvious edges by 3$\sigma$ threshold filtering, and reconstruct the time-frequency dynamic spectrum by using the two-dimensional wavelet algorithm. This algorithm tends to retain more data than the traditional frequency channel zapping methods. In \textcolor{blue}{Extended Fig}. \ref{fig:RFIs}, we demonstrate the effect of RFI removal and the corresponding RFI-mitigation data (panel A and B). It is obvious from the frequency bandpass that nearly all narrow-band RFIs have been masked in RFI-mitigation data (panel C), and the histogram is consistent with a Gaussian white noise distribution (panel D, the value of Chi-square is 5$\%$).

\begin{figure}
\centering
\includegraphics[width=.9\textwidth]{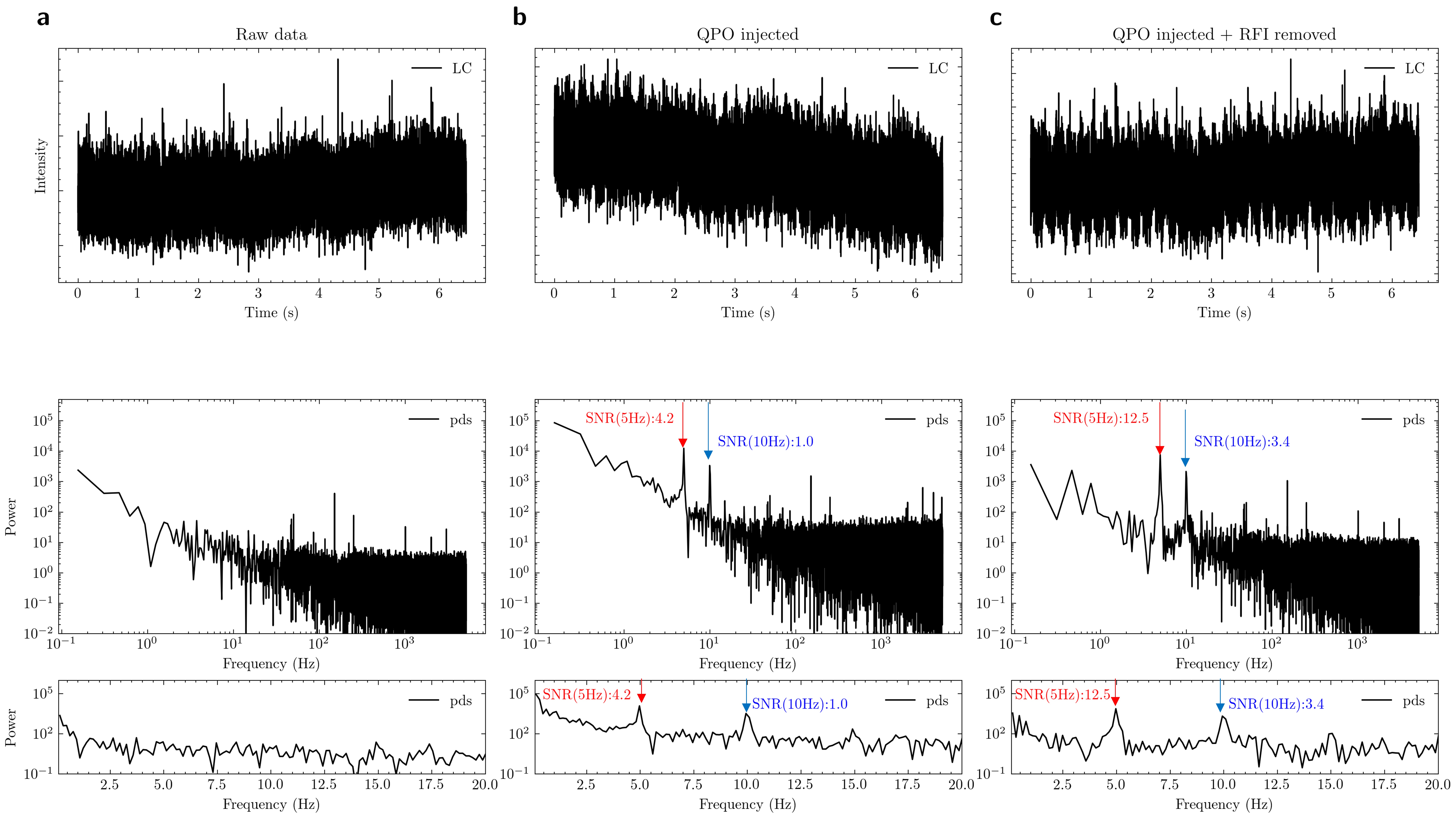}
\caption{{\bf Example of QPO signals and RFI removing simulations.} Upper panels show frequency-averaged light curves, subplots {\textbf{a/b/c}} are the raw data of sky background monitoring, simulated injection of 5 Hz and 10 Hz QPOs in the broad bands, and the light curve after removing all RFIs, respectively. Bottom panels show the corresponding Fourier power spectra in logarithmic and linear coordinates. Subplot {\bf c} demonstrates the apparent increase in the significance of the detected QPO signals, i.e. from 4.2 $\sigma$ to 12.5 $\sigma$ for 5 Hz; from 1 $\sigma$ to 3.4 $\sigma$ for 10 Hz, compared with that in Subplot {\bf b}. The simulations demonstrate that our RFI removing processes can efficiently reduce the narrow-band RFIs and keep the broad band astrophysical signals. }
\label{fig:Noisedoffsource}
\end{figure}

To ensure that the RFI removal algorithm does not block the detection of QPO-like signals, we designed an experiment to simulate the injection of 5 Hz and 10 Hz temporal modulated broadband signals into the real FAST data to evaluate the effect of the two-dimensional wavelet algorithm. The combined effect of 5 Hz and 10 Hz temporal intensity injection and RFI events, particularly the satellite bands around 1.2 GHz, are shown in \textcolor{blue}{Extended Fig}. \ref{fig:Noisedoffsource} (panel B). As a comparison, the result of RFI-mitigation data (panel C) shows that not only the two-dimensional wavelet algorithm does not block the detection of the corresponding periodic signal, but it also increases the significance of the detected signal (e.g., from 4.2 $\sigma$ to 12.5 $\sigma$ for 5 Hz; from 1 $\sigma$ to 3.4 $\sigma$ for 10 Hz).

\begin{figure}
\centering
\includegraphics[width=1.0\textwidth]{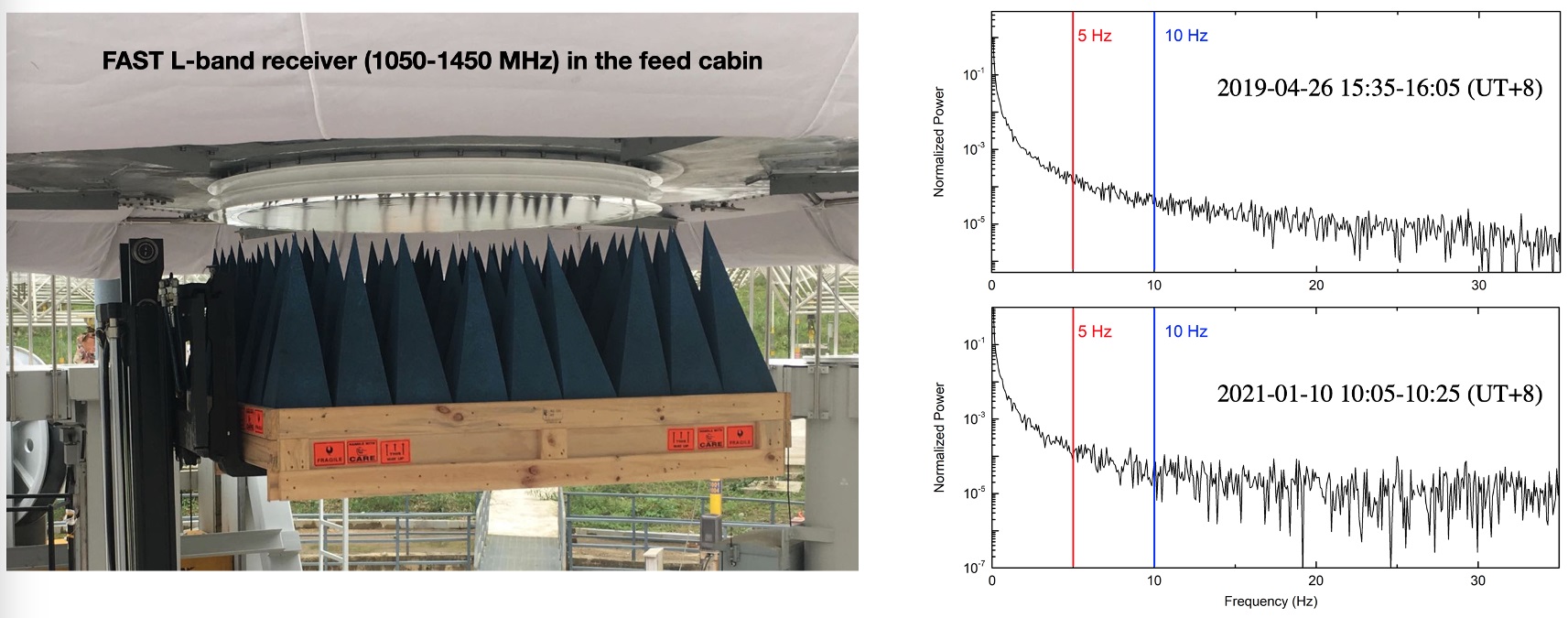}
\caption{{\bf The 19-beam receiver performance.} {\bf Left panel:} an absorber is used to cover the receiver feed opening during noise tests. {\bf Right panel:} no-detection of 5-Hz or 10-Hz apparent peaks from FAST receiver itself on two of frequency-averaged light-curve time segments.}
\label{fig:receiver}
\end{figure}

We also worked with the FAST EMC technical team to carefully examine the noise signals covering the whole bandpass to check possible QPO-related effects from the receiver or backend instrumentation. The FAST EMC technical team regularly conducts noise tests on the L-band receiver performance \cite{jiang2020}. During each of the noise test, an absorber is used to cover the receiver feed opening (\textcolor{blue}{Extended Fig}. \ref{fig:receiver}, left panel), i.e. the receiver will not receive any of external signals during the time. We re-analyzed the switch-by-switch RFI exclusion experimental data on 2019-04-26 15:35-16:05 (UT+8) and 2021-01-10 10:05-10:25 (UT+8), as demonstrated in the corresponding power spectrum for each of the two frequency-averaged light-curve segments in \textcolor{blue}{Extended Fig}. \ref{fig:receiver} (right panel) assuming a broadband signal (a few hundred MHz). No detection of 5-Hz or 10-Hz signals within the full of frequency range (1050-1450 MHz) was made. The horizontal and vertical axes show the Fourier frequency and the intensity in arbitrary units of the power spectrum, while the vertical red and blue lines are the locations of 5Hz and 10Hz, respectively.  Throughout the observations, no QPO-liked signal (in particular for 5-Hz) with an apparent intensity above 3$\sigma$ was detected.

{\bf Flux calibration} \\
The system temperature of the telescope is a function of zenith angle,
\begin{equation}\label{Tsys}
\begin{split}
    T_{sys} = P_0\cdot \arctan(\sqrt{1 + \theta_{ZA}^n} - P_1) + P_2,
\end{split}
\end{equation}
where $\theta_{ZA}$ is the zenith angle, $P_0$, $P_1$, $P_2$ and $n$ are parameters, and their values vary with different frequencies. Table~4 and Fig~12 in \cite{jiang2020} show the relationships between $T_{sys}$ and $\theta_{ZA}$ for different beams. The temperature of the source is,
\begin{equation}
\begin{split}
    T_{src}(t) =T_{cal}\cdot\frac{ON}{CALON-ON} - T_{sys}(t),
\end{split}
\end{equation}
where $T_{cal}$ is the temperature of the injected reference signal from the noise diode, $CALON$ and $ON$ are the intensity values with the injected signal switched on and off for the calibration scans before and after the tracking observations, and $T_{sys}(t)$ is the time-dependent system temperature. Then the flux density can be calculated by the following formula,
\begin{equation}
\begin{split}
    Flux(t) = \frac{Track(t)}{ON}\cdot T_{src}(t)\cdot\frac{1}{G},
\end{split}
\end{equation}
where $Track(t)$ is the intensity values during the tracking observations, $G=\eta G_0$ differs from the measured gain $G_0$ = 25.6 K/Jy by a factory $\eta$, is the full gain of FAST in the sky coverage, $\eta$ is the aperture efficiency.

\begin{figure}
\centering
\includegraphics[width=.7\textwidth]{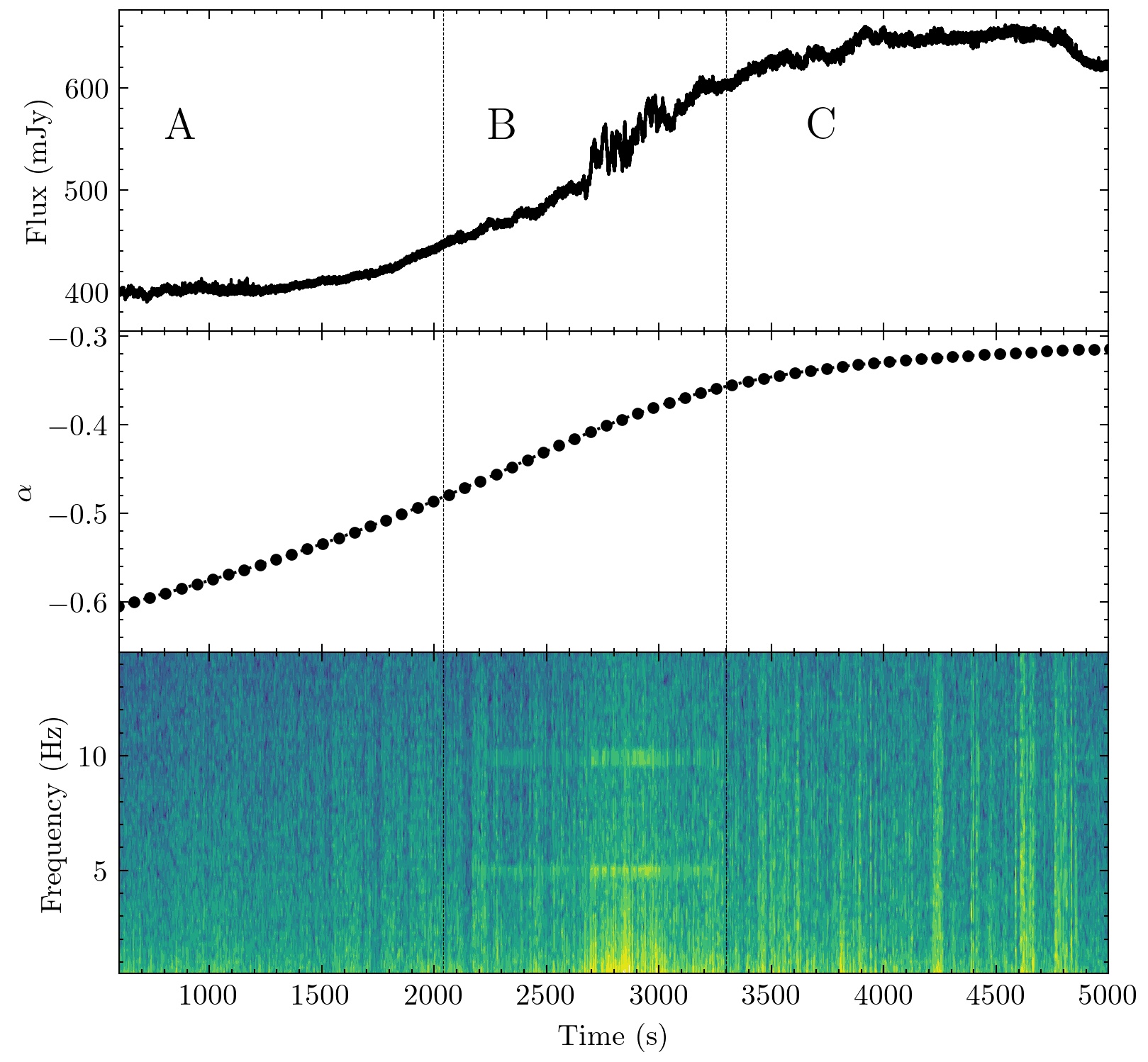}
\caption{{\bf Light curve and dynamic power spectrum with less channels.} Lightcurve, spectral index evolution and dynamic PDS of the radio flux which is calibrated with directly removing RFI peaks from the channels 1400 to 2380. There are 2700 channels left after channel cutting and RFI removing. QPO signals at $\sim 5$ Hz are fainter compared to the case with about 3400 channels. The Epochs A, B and C have the same definition in Fig. \ref{fig:lc-pol}, with the time record starting from January 25 2021 01:35:00 (UTC). The index $\alpha$ varies from $-0.6$ to $-0.3$ during the Epochs. }
\label{fig:1300-2400}
\end{figure}

Meanwhile, there are some RFI broad peaks around channels from 1400 -- 2380 (see \textcolor{blue}{Extended Fig}. \ref{fig:RFIs}) which are conspicuous. For a test for the very clean data base, we did another calibration and dynamic PDS calculation for lightcurve with channels from 1400 -- 2380 removed directly, and the results are shown in \textcolor{blue}{Extended Fig}. \ref{fig:1300-2400}. The QPO signals are still recognizable, but a little weaker, which would be due to the reduce of observed channels leading to the lower signal-noise ratio. In addition, we also derive the spectral index in the band 1.05 -- 1.45 GHz defined by $\alpha=\Delta \log {S_{\nu}}/\Delta \log\nu$. The spectral index evolution versus time is shown in \textcolor{blue}{Extended Fig}. \ref{fig:1300-2400}, $\alpha$ varies from $-0.6$ to $-0.3$ during the increasing of the flux. The change of the radio spectrum generally occurs prior to the ejection event \cite{klein2002,fender1999}.

{\bf Polarization calibration} \\
The original {\tt PSRFITS} files of FAST observation contain the polarization components which are recorded as the AABBCRCI form, where AA and BB are the direct products of two channels, CR and CI are the real and imaginary parts of the cross product of two channels respectively. We use {\tt DSPSR} \cite{straten2011} and {\tt PSRCHIVE}  \cite{hotan2004} to fold the original files with their duration as folding periods to produce the four time series. Each of the four time series is full of 4096 frequency channels and 128 time subints, so it is necessary to eliminate the RFIs manually for each time series.

We calculate the stokes parameters from the four recorded channels, named $I_x^2$, $I_y^2$, CR and CI respectively \cite{straten2010}, where CR and CI are the real and imaginary parts of the cross product of two channels $I_x*I_y$. Generally, for normal devices, feeds are never perfect, and there are two quantities that should be considered and calibrated, i.e., relative gain of electronic system and phase differences between two channels. In this paper, we named the two quantities as leakage and phase error. And stokes parameters with subscript $obs$ and $true$ refer to polarization components before and after calibration which normally should take the Mueller matrix \cite{heiles2001a,heiles2001b} of the equipment into account respectively, then considering the leakage between two channels, for the linearly polarized signal from the diode, we find
\begin{equation}
\begin{array}{l}
    I_{obs}^{\prime} = I_{true}^{\prime} + f*Q_{true}^{\prime}\\
    Q_{obs}^{\prime} = Q_{true}^{\prime} + f*I_{true}^{\prime},\\
\end{array}
\end{equation}
where $\prime$ means the injected reference signal, and the leakage $f=\frac{Q_{obs}^{\prime}}{I_{obs}^{\prime}}$.
We calibrated the phase error as $\delta_{er}=\frac{1}{2}\arctan\frac{V_{obs}^{\prime}}{U_{obs}^{\prime}}$, then removing the error from orientation, we can get the true values of the four stokes parameters:
\begin{equation}
\begin{aligned}
    I_{true}&=\frac{I_{obs}-f*Q_{obs}}{1-f^2}\\
    Q_{true}&=\frac{Q_{obs}-f*I_{obs}}{1-f^2}\\
    U_{true}&=P\cos[2(\delta-\delta_{er})]\\
            &=P\cos2\delta \cos2\delta_{er}
             +P\sin2\delta \sin2\delta_{er}\\
    V_{true}&=P\sin[2(\delta-\delta_{er})]\\
            &=P\sin2\delta \cos2\delta_{er}
             +P\cos2\delta \sin2\delta_{er},
\end{aligned}
\end{equation}
where $P=\sqrt{U_{obs}^2+V_{obs}^2}$. Finally, the degrees of linear and circular polarization, and polarization position angle are calculated as
\begin{equation}
\begin{aligned}
    LP&=\frac{L}{I_{true}}=\frac{\sqrt{Q_{true}^2+U_{true}^2}}{I}\\
    CP&=\frac{V_{true}}{I_{true}}\\
    PA&=\frac{1}{2}\arctan\frac{U_{true}}{Q_{true}}.
\end{aligned}
\end{equation}

\section{Dispersion Measure Test}

\begin{figure}
\centering
\includegraphics[width=.65\textwidth]{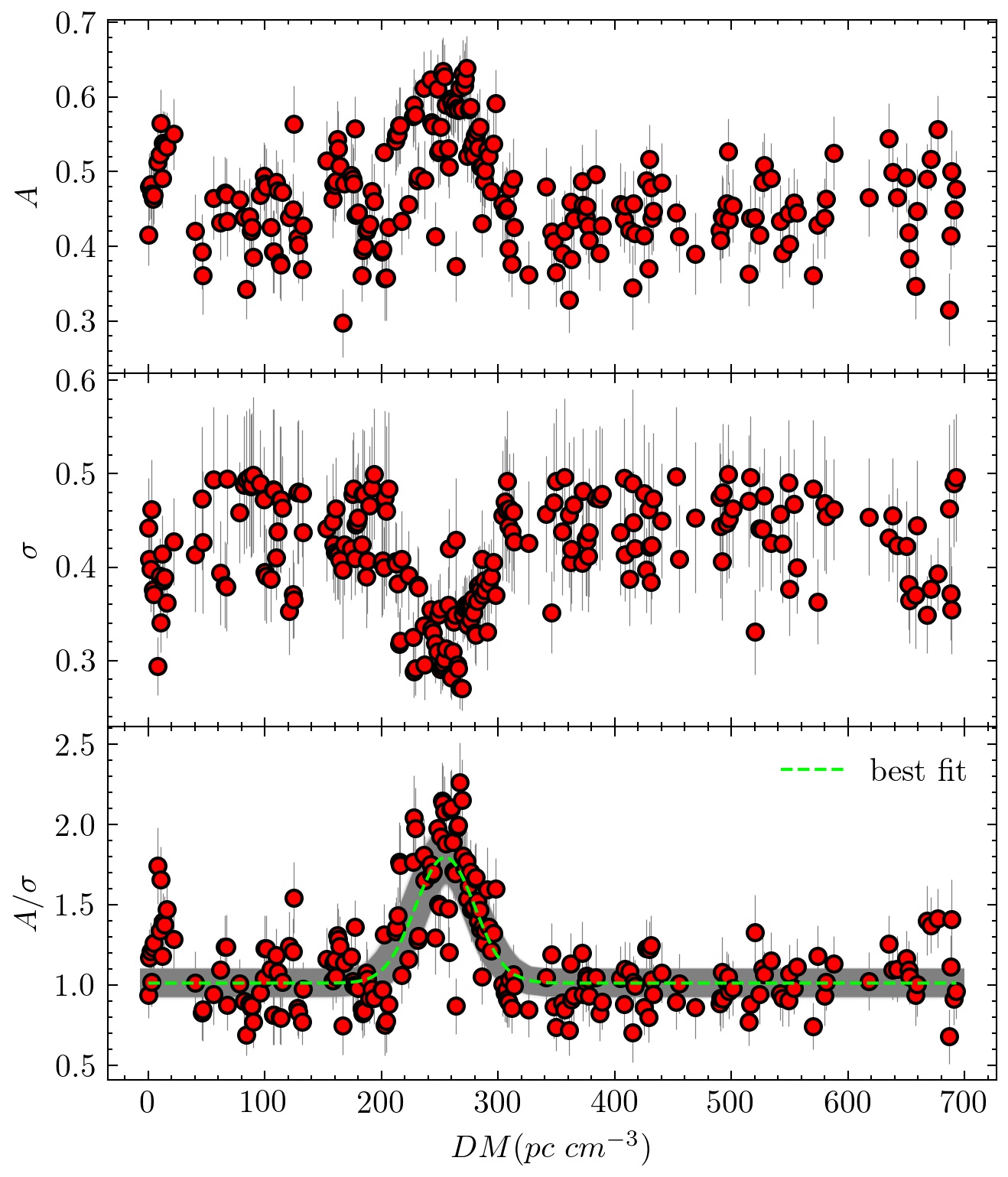}
\caption{{\bf The DM value of QPO signals.} Amplitude ($A$) and full width at half maximum ($\sigma$) of the fitting Lorentzians for the folded curves evolve with DM. The peak of $A/\sigma$ is located at DM$\sim 255 \pm 25\ \rm pc \ cm^{-3}$, which is fitted via a Gaussian function (green dashed line), would indicate the possible dispersion measure of GRS 1915+105.} \label{fig:Significance}
\end{figure}

The dispersion measure (DM) is the quality in a pulsar, representing integrated column density of free electrons between an observer and the pulsar \cite{Davidson1969}, then observationally as a broadening of an otherwise sharp pulse when a pulsar is observed over a finite bandwidth. Unfortunately, the very broad, and sine-like profile of transient 5-Hz QPO compared to the pulses in pulsars makes it difficult to directly detect the DM of the source from the delay of time of arrival (TOA) between different frequencies of the QPO. In this work, we used the  {\tt prepfold} to fold the data during the Epoch B in January 2021 at the period of the QPO (0.196 s) at different DM values. Here we take the DM ranges from 0 -- 700 $pc\ cm^{-3}$ and set the step of 1 $pc\ cm^{-3}$. And then, we fit each folded profiles via Lorentzians. The fitting parameters, e.g. amplitude ($A$) and full width at half maximum ($\sigma$) are supposed to reach the maximum and minimum respectively around the true DM value. In \textcolor{blue}{Extended Fig}. \ref{fig:Significance}, the evolution of $A$, $\sigma$ and $A/\sigma$ with DM are plotted, the peak of $A/\sigma$ centering around DM$\sim 255\ \rm pc\ cm^{-3}$ with a standard deviation of $25\ \rm pc\ cm^{-3}$ ($68\%$ confidence level) indicates the possible dispersion measure of the source.

\section{Dynamical power spectrum}

\begin{figure}
\centering
\includegraphics[width=.60\textwidth]{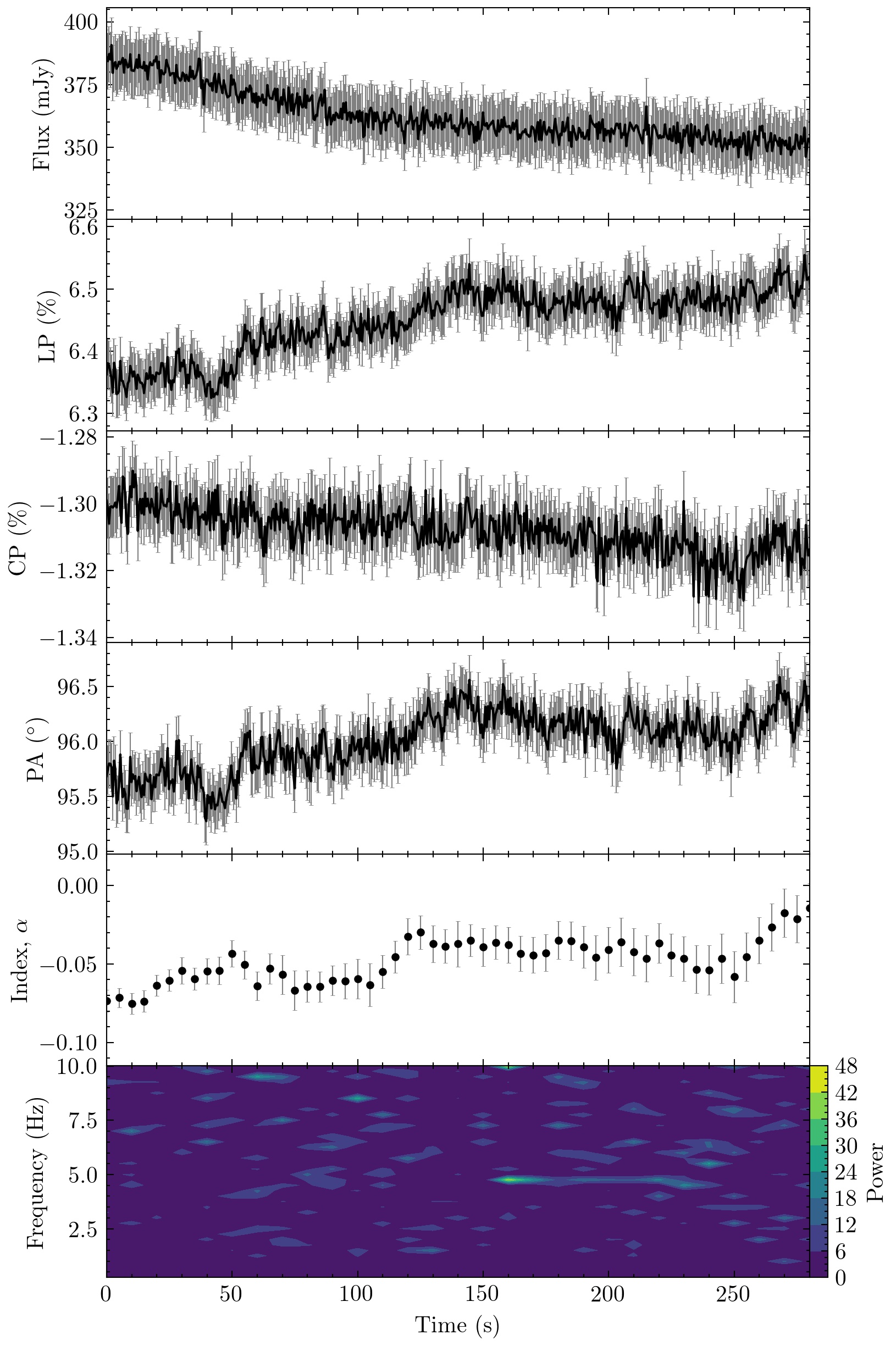}
\caption{{\bf Light curves and dynamical power spectrum during the QPO phase in 2022.} The light curves of total intensity flux density, LP, CP, PA, spectral index $\alpha$ and dynamic PDS with FAST observations from 2022-06-16:17:42:40 to 2021-06-16:17:47:30 (UTC). The transient QPO at $\sim 5$ Hz lasting about 80 seconds was detected. During the event, the radio flux was steady at a level around 350 mJy; LP was around $6.5\%$ and increased slightly during the observations; CP was measured at $\sim -1.3\%$, and  the PA was around 96$^\circ$. The spectral index $\alpha$ also evolved from $-0.08$ to $-0.01$. All error bars are given at the 1$\sigma$ level. }
\label{fig:lc2022}
\end{figure}

The main aim of the timing analysis here is to search for the sub-second quasi-periodic oscillations (QPOs) in radio flux light curves. QPOs are generally studied in the Fourier domain and showing up in the power density spectrum as narrow peaks. We used {\tt Numpy.fft.fft} and {\tt Stingray} in {\tt Python} packages to perform the power density spectrum (PDS) analysis, including the production and fitting of the PDS. We calculated the power density spectrum for every 4 second data set of the calibrated flux time series, and then we arranged them in chronological order to get a set of spectra which is called as the dynamical power spectrum (see Fig. \ref{fig:lc-pol} \& \textcolor{blue}{Extended Fig}. \ref{fig:1300-2400}) for the whole observational time series on January 25 2021. Based on the QPO signal evolution behavior, the observational time series can be divided into three time regimes: the pre-QPO regime (Epoch A), QPO regime (Epoch B), and post-QPO regime (Epoch C). With the same methods, we also derived the evolution of the radio flux and polarization for the second QPO event on June 16 2022, and presented the dynamical power spectrum of the flux light curves in  \textcolor{blue}{Extended Fig}. \ref{fig:lc2022}. The QPO around 5 Hz lasted for about 80 seconds, which was also weaker than that detected in January 2021.

In Fig. \ref{fig:event2022}{\bf a}, we displayed the PDSs of the three light curves selected from the three epochs (A, B and C) on January 25 2021. The plots of the three PDSs show two unambiguous and significant peaks around 5 and 10 Hz only appearing during Epoch B. There are no QPO signals during Epochs A and C. In addition, for comparison, we also showed the PDSs of the three light curves selected from the central beam and other two beams with the FAST observations performed on June 16 2022 in Fig. \ref{fig:event2022}{\bf b}. The 5-Hz QPO was only reported in the central beam towards the source. No signals were detected in other beams towards the background sky regions. Meanwhile, in Fig. \ref{fig:event2022}, we denote a significance level of 3 $\sigma$ using a light curve simulation algorithm \cite{Timmer1995} to represent the criterion for QPO detection. For the significance level computation, we simulated 20000 light curves with power-law distributed noises appropriate for our data and re-sampled these light curves to ensure the resolution that matched our observation data.

\section{Wavelet analysis results}

The dynamical power density spectra (PDS) technique has the limit of the time resolution at the time intervals (several seconds) used for the time domain analysis. However, wavelet analysis method can provide accurate time-frequency space information with very high time resolution, thus can be used to study the detailed variation characteristics
of the periodic or quasi-periodic signals over time \cite{torr1998}, which also has been applied to timing analysis of X-ray light curves in the X-ray binaries to discover the transient QPOs \cite{ding2021,chen2022,chen2022b}.

We also used the wavelet analysis method to test the QPO signals we detected in dynamical PDS, which can approach the time evolution and variation of the signal. In wavelet analysis, we have taken the red noise into account by calculating the correlation functions of the time series, and a simple model to compute red noise is the univariate lag-1 autoregressive process, so that we estimate the red noise from $(\alpha_1 + \sqrt{\alpha_2})/2$ where the $\alpha_1$ and $\alpha_2$ are the lag-1 and lag-2 autocorrelations of the time series. Based on the chi-squared distribution, if a power in the wavelet power spectrum is above the $95\%$ confidence level compared with the background spectrum, then it can be considered as a true signal.

The wavelet power spectrum with the time shows that the QPOs have the fine structure evolution both in the time and frequency domains (see the example of the wavelet spectrum in Fig. \ref{fig:wavefold}). The 5-Hz QPO signal is stronger over the time than the 10-Hz QPO, in addition, 5-Hz QPO can be detected in most observational time intervals during the QPO regime, while the 10-Hz QPO signal distributes sparsely. And then based on the wavelet power spectrum, we can clearly identify the time regimes when the 5-Hz QPO is only detected, and when both 5-Hz and 10-Hz QPOs appear. Thus, we fold the light curves at the 0.2-s period to create the pulse profiles of the flux density for the 5-Hz regime, which shows the single-peak broad pulse profile. In addition, to check the variation patterns of the polarization with the pulse profiles of the QPOs, we also fold the light curves of different polarization components (i.e., LP, CP and PA) at the 0.2-s period of the QPO for the 5-Hz regime (see Fig. \ref{fig:wavefold}).

\begin{figure}
\centering
\includegraphics[trim=0 0 40 0,width=.65\textwidth]{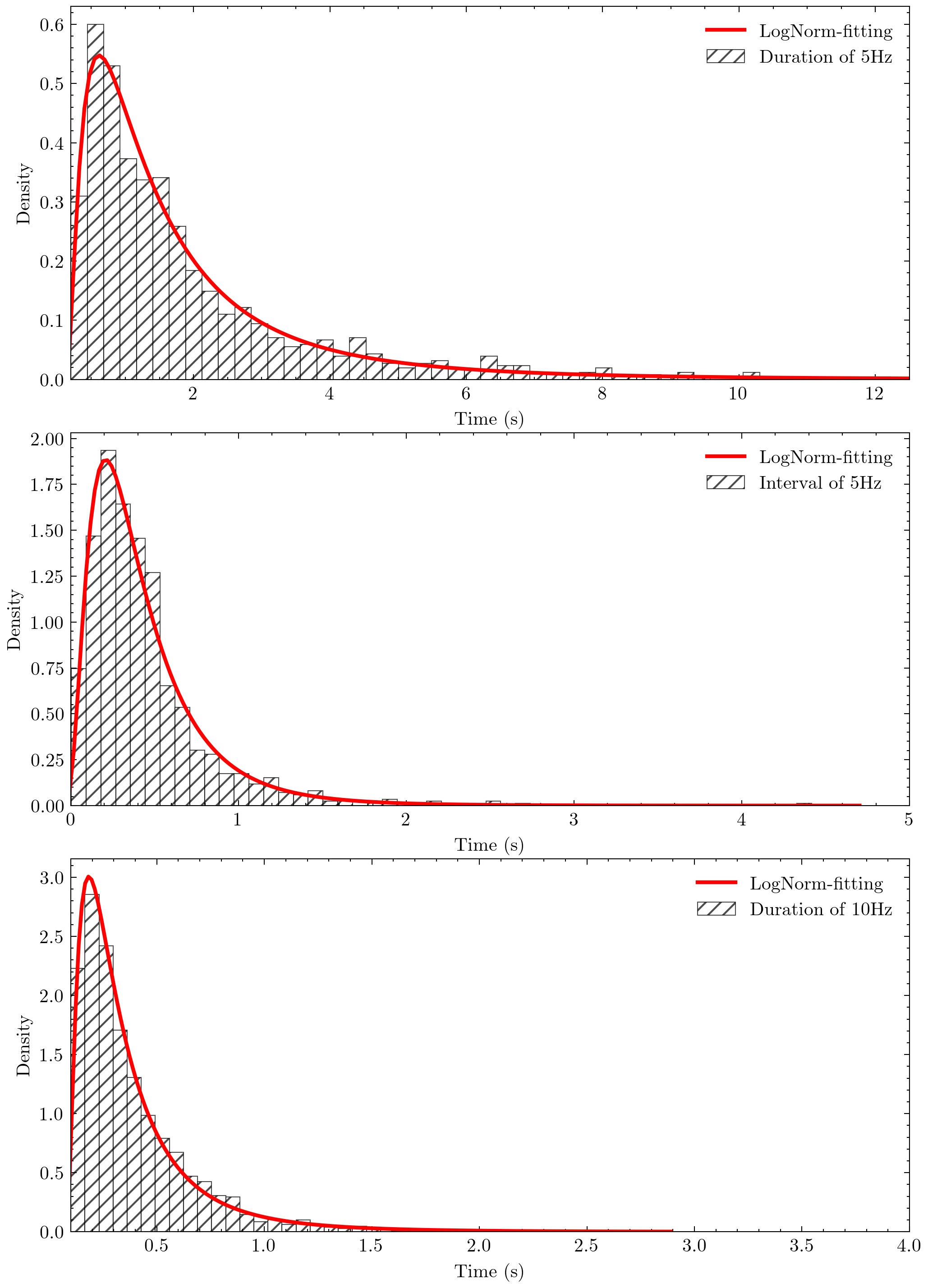}
\caption{{\bf The distributions for three time scales observed in Epoch B.} The duration distribution of the 5-Hz QPOs (top), separate interval distribution for two neighbour 5-Hz QPOs (middle), and duration distribution of the 10-Hz QPOs (bottom), and the red lines are the best fitting curves with the log-normal distribution. }
\label{fig:distribution}
\end{figure}

With the time-frequency space information provided by wavelet analysis, the variance of power with time and frequency can be easily identified, so as to distinguish the time intervals with QPOs and non-detection of QPOs. The radio variation properties should be related to the jet dynamics, then the short time scale evolution of two QPO signals provides the probe of the characteristic time scales of jet production and dynamics. Thus we make statistics on time intervals of the QPO signals in Epoch B (see \textcolor{blue}{Extended Fig}. \ref{fig:distribution}): the duration distribution of the 5-Hz QPOs, separate interval distribution for two neighbour 5-Hz QPOs, the duration of 10-Hz signal. The 5-Hz signal is strong and can be detected in most time. While, the 10-Hz signal is weaker and the QPO feature lasts for only about one second or sub-seconds and there are gaps lasting for several to tens of seconds without a feature.

We also use the logarithmic normal function to fit the distributions to determine the peak values of three typical time intervals, which will probe the characteristic dynamical time scales of jets near the BH. The duration of the 5-Hz QPO distributes in the broad time scales from 0.3 -- 12 s, and the peak around 0.7 s. This characteristic time scale would be connected to the typical emission size of the QPO emission source in the jet ($\sim c\tau$). Though the 5-QPO signal is the dominant component in the power spectrum, the signal would be also not continuous, and in some time intervals of Epoch B, no QPO signal can be detected.

\section{Other radio QPOs in BH accretion systems }
Long-period radio QPOs with the periods from about one hundred days to several years have been reported in some radio loud active galactic nuclei (AGNs), specially blazars \cite{zhang2021,Ren2021,raiter2001,Bhatta2017}. These radio QPOs generally last for about several to twenty cycles, which likely reflect the special dynamics of relativistic jets powered by supermassive black holes (SMBHs) in AGNs. Radio oscillations with a period of $\sim$ 15 hours was also found in a gamma-ray X-ray binary LS I$+61^\circ$303 \cite{Jaron2017}, which only had two or three QPO cycles. In addition,  slow radio oscillations in the period range of the 20-50 minutes were detected in GRS 1915+105 \cite{pooley1997,rodri1997,klein2002}.

There have been a few physical models suggested to interpret these QPOs. For stellar mass BH systems, e.g., GRS 1915+105, the half-hour radio periodic oscillations may be connected to the X-ray oscillations with the similar periods, while in LS I$+61^\circ$303, it was suggested that the radio QPOs could result from multiple shocks in a jet \cite{Jaron2017}. The infrared QPOs around 0.1 Hz reported in a microqusar GX 339-4 are attributed to the jet precession \cite{Malzac2018}. In the framework of AGNs, radio QPO models are diverse. The year-long QPOs are generally considered to be the indicator of the orbital motion of binary SMBH systems. Helical structures in magnetic fields and plasma trajectory are expected in magnetically dominated jets \cite{chen2021}, so helical motion of blobs or shocks in relativistic jets have been incorporated to interpret periods around hundreds of days in radio, optical or gamma-ray bands in blazars \cite{zhou2018,Sarkar2021}. Recently, the optical and gamma-ray periods around 0.6 day in BL Lacertae were suggested to originate from kink instability in relativistic jets \cite{Jorstad2022}.

\section{X-ray monitoring of GRS 1915+105}
\begin{figure}
\centering
\includegraphics[trim=0 0 40 0,width=.8\textwidth]{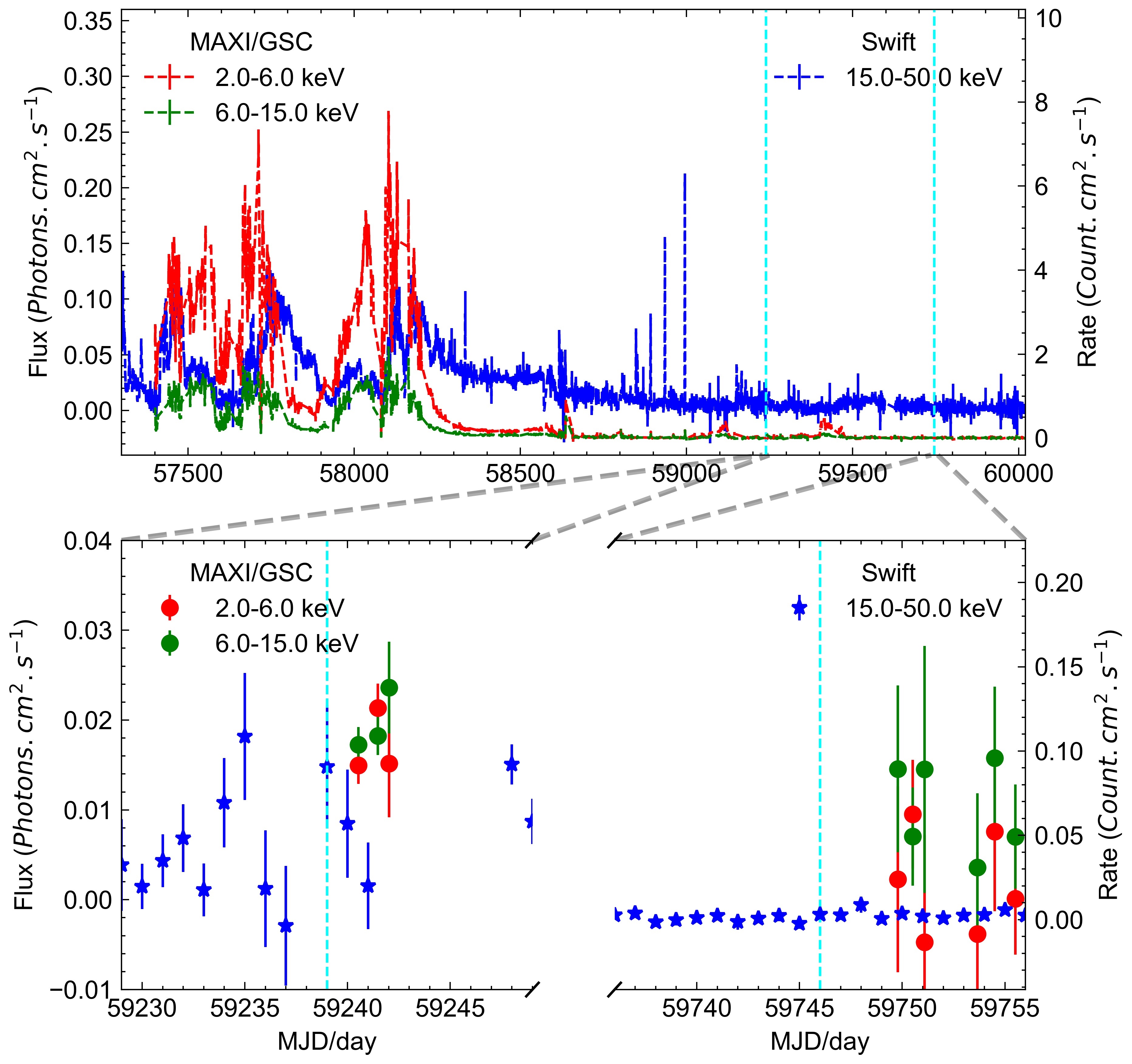}
\caption{{\bf X-ray monitoring during the radio QPOs}. {\bf Top:} The X-ray light curves of GRS 1915+105 from 2016 - 2023 based on the SWIFT and MAXI long-term monitor observations. {\bf Bottom:} The zoom-in version of the light curves around January 25 2021 and June 16 2022. The vertical dashed lines show the time of our FAST observations. }
\label{fig:lc_xray}
\end{figure}

We have checked the X-ray light curves by monitoring the source based on Swift and MAXI/GSC from 2016 - 2021, which are displayed in \textcolor{blue}{Extended Fig}. \ref{fig:lc_xray}. Swift/BAT covers the energy band of $15- 50$ keV, and MAXI reports the count rates of two energy bands: 2 -- 6 keV and 6 -- 15 keV. As expected, GRS 1915+105 is the strongly variable X-ray source, and shows flares in the historic records. Since 2018, GRS 1915+105 unexpectedly started a peculiar low-luminosity state that is an order of magnitude dimmer than the previous states, with greater hardness ratio in X-rays \cite{Negoro2018,Neilsen2020,Miller2020,Koljonen2020,Koljonen2021,Ratheesh2021} (also see \textcolor{blue}{Extended Fig}. \ref{fig:lc_xray}). Even though intrinsic dimming is possible, detailed X-ray spectral analyses suggested that the source may have entered an obscured state with the strong absorption (by disk winds or torus in the outer disk part) along the observer's sight \cite{Neilsen2020,Miller2020,Koljonen2020,Ratheesh2021,Balakrishnan2021} due to a large inclination angle of $\sim 60^\circ$ \cite{reid2014}. This interpretation was supported by the detection of X-ray flares which are not strongly affected by obscuration. The observations of frequent radio flares due to the episodic jet emissions \cite{Motta2021} are also consistent with this scenario. During the FAST observations on 2021-01-25 and 2022-06-16, the X-ray flux was weak based on both Swift and MAXI observations. The radio oscillations revealed by our FAST observations suggests that presently GRS 1915+105 may still have a high accretion rate to power transient relativistic jets. This adds further support to the suggested strong obscuration in X-rays.

\section{Comparison with other radio observations}
The radio flux and LP values from relativistic jets are highly variable, and change significantly in different radio bands and different epochs\cite{klein2002,fender2002,fender1999,Mjones2005,rusht2010}. Here we would briefly compare our radio results on GRS 1915+105 with other observations. VLA, VLBA, MeerKat, MERLIN and other radio telescopes have monitored GRS 1915+105 in different radio bands. MeerKat reported a radio flux of $\sim 100-900$ mJy during the radio flares after 2018 \cite{Motta2021}, which is consistent with the radio flux of the present observations. In addition, MERLIN has the similar observational waveband around 1.2 GHz, reporting a LP from $\sim (1-24) \%$  \cite{Mjones2005,rusht2010}. Our FAST observation on 2021-01-25 has a LP ranging in $\sim (25-31)\%$, which is a little higher than or still approach the previously reported values. This value is also physically reasonable. For a magnetically dominated jet with an ordered magnetic field configuration, in the optically thin emission ($\alpha<0$, e.g., the case of radio flares) the maximum LP can be as high as 70 \% \cite{longair1994,curran2014}. For the observations of FAST on 2021-01-25, the radio flare has $\alpha\sim -0.5$, the LP could be as high as $\sim 30\%$ for a large-scale magnetic field along the jet, consistent with the FAST result. The somewhat smaller LP in earlier observations may be a result of observing different episodes of jet injection. As the jet propagates, the LP degree may degrade because the dissipation of the ordered magnetic field in the emission region. Our observation might have caught the early phase of a freshly injected jet that has a higher LP degree. In general, the linear polarization of the present observations is still similar to the previous observations and other BH systems\cite{Mjones2005,rusht2010,curran2014}, which increases with decreasing randomization of the magnetic field within the jet component\cite{fender1999}.

\section*{Data availability statement}
All FAST data are available from the FAST user website, \url{http://fast.bao.ac.cn}.

\section*{Code availability}
PSRCHIVE (\url{http://psrchive.sourceforge.net})\\
DSPSR (\url{http://dspsr.sourceforge.net})\\
PRESTO (\url{https://github.com/scottransom/presto})\\

\section*{Acknowledgements}
This work is supported by the National Key Research and Development Program of China (Grants No. 2021YFA0718500, 2021YFA0718503), the NSFC (12133007, U1838103, U2031117), the Youth Innovation Promotion Association CAS (id.2021055), CAS Project for Young Scientists in Basic Research (grant YSBR-006) and the Cultivation Project for FAST Scientific Payoff and Research Achievement of CAMS-CAS.

\section*{Author contributions statement}
W.W. as the PI of the FAST observations led the data analysis and wrote the paper. P.T. and P.Z. made the data analysis, P.W., X.S., J.L. and Z.Z. provided the help of the radio data analysis and software. W.W., B.Z., Z.D., F.Y., S.Z., Q.L. and X.W. constructed the scientific interpretation to the data and B.Z. contributed to paper writing. P.W., P.J., D.L., H.Z., Z.P. and H.G. aided with the FAST observations. J.C., X.C. and N.S. provided the X-ray data. All authors have reviewed the present results and manuscript.

\section*{Competing interests}
The authors declare no competing interests.

{}

\end{document}